# Predicting economic growth with classical physics and human biology

Hans G. Danielmeyer and Thomas Martinetz

Institut für Neuro- und Bioinformatik, Universität zu Lübeck, Germany
www.inb.uni-luebeck.de

**Abstract**
We collect and analyze the data for working time, life expectancy, and the pair output and infrastructure of industrializing nations. During S-functional recovery from disaster the pair's time shifts yield 25 years for the infrastructure's physical lifetime. At G7 level the per capita outputs converge and the time shifts identify a heritable quantity with a reaction time of 62 years. It seems to control demand and the spare time required for enjoying G7 affluence. The sum of spare and working time is fixed by the universal flow of time. This yields analytic solutions for equilibrium, recovery, and long-term evolution for all six variables with biologically stabilized parameters.

The first obstacle to predicting economic growth is having no information on demand beyond existential needs. It can neither be measured with money nor added to the national level like the costs of production because user values are partially emotional and overlap for user time. The exponential function is generally used for approximating growth. The annual output $y(t)$ called gross domestic product (GDP) per capita is described with mathematical products of the inputs work $w(t)$ and physical capital $k(t)$, the technical infrastructure.

For the G7 the basic needs are satisfied. Advertising and short-term forecasting with dozens of indicators have become big business. But socioeconomic policy is limited to correcting mistakes until the demands for $w(t)$ and investment for $k(t)$ are quantitatively understood.

The deadlock seems hopeless, but physics has a successful tradition of using symmetries for discovering hidden quantities on the other side. The resulting theory should be as simple as possible, reproduce all available data without adjustable parameters, yield continuous equilibrium between demand and supply in times of peace, and predict medium and long-term growth paths for the complete set of variables. Considering the industrial society's political and financial disasters from 1861 to 2008 the actual existence of such a theory may be its only support. But the following theory agrees with all available data. The early data were taken from B. R. Mitchell's International Historical Statistics (Stockton Press Series 1755-1988). Recent data are 10-year averages of the data reported in the national statistical yearbooks.

The following figures show six new facts. The lower part in figure 1 shows the recoveries of the leading nation's inflation corrected GDP $y(t)$ per capita. They converge into a collective envelope $a(t)$ named "industrial evolution" [1]. The Great Depression stopped the first recovery of the USA from the Civil War. The second began after the peak in 1942 with the end of World War II shortly before the fast recoveries of West Germany and Japan. South Korea's convergence after the Korean War is past halftime. The upper part shows the parallelism between $a(t)$ and the increase of the pioneering nation's mean unisex life expectancy $L(t)$ [2]. The six lowest values are due to initially higher child mortality. Figure 2 shows the time shifts between the normalized national paths of $y(t)$ and $k(t)$ as well as an anti-correlation of the working time $w(t)$ with $a(t)$ and $L(t)$.

Both figures show only S-functions. They can be fitted to real processes, but reliable predictions cannot be based on extrapolating data by fitting functions and/or parameters. We complete the set of variables and parameters, quantify all interactions, find S-functions as analytic solutions, and measure the system's parameters with the entire pattern of normalized S-functions. For irreversible processes they display physical lifetimes of storable quantities with time shifts like Sinus-functions display them for periodic processes with phase shifts. The lower bars in figure 2 show the time shifts $\Delta\tau = (1/\beta)Log(1+\beta G)$ where $0.05 \leq \beta \leq 0.10$ are the initial national growth rates of recovery and $G = 25$ years is the effective lifetime of $k(t)$. These bars are the first measurements of $G$. The upper bars show always the same time shift $\Delta T = E Log(1 + G/E) = 21$ years. The corresponding reaction time of $E = 62$ years identifies a second, so far hidden storable quantity. We name it human capacity $h(t)$. When it combines inherited with educated capacities it has the potential for controlling $L(t)$ and $a(t)$ simultaneously.

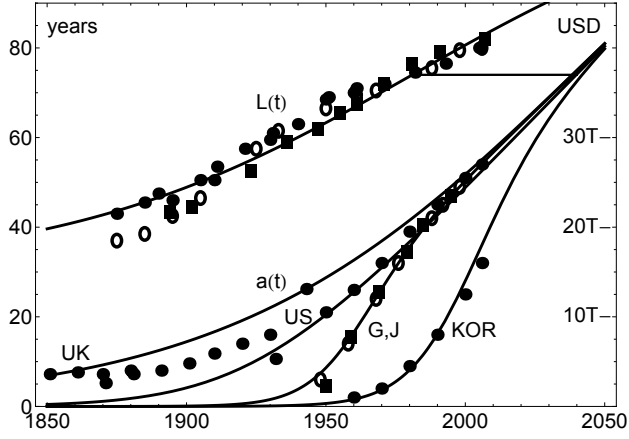

*Figure 1: Real GDP per capita (right hand scale in US $ 1000 of 1991) and unisex life expectancy (left hand scale) for the UK and the USA (points), Germany (circles), and Japan (squares) compared with theory (plots). South Korea can catch up with the G7 in 2040. L(t) and a(t) are from [2].*

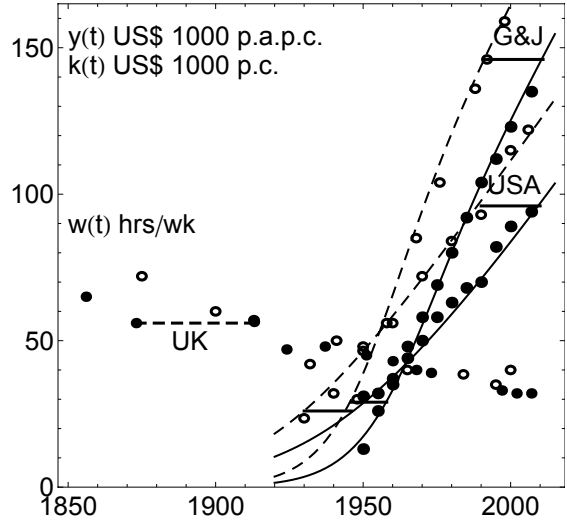

*Figure 2: The time shifts (4 bars) of physical capital k(t) (points) with respect to the normalized GDPs (circles) compared with theory (plots) and the working time in Germany (circles) and the UK (points). The UK's dashed stagnation is also observed in the GDP of figure 1.*

The time shifts follow from generating $k(t)$ with three terms: the maintenance of the accumulated value $k(t)$ with a part $0.08 \leq \mu \leq 0.26$ of the GDP given by $k/G = \mu y$, and for constant $G$ the annual investment $\dot{k} = \mu G \dot{y} + \dot{\mu} G y$ with two parts. The total national support

$$\dot{k} + k/G = \mu(1 + G\dot{y}/y)y + \dot{\mu}Gy \equiv \bar{\mu}y + \dot{\mu}Gy \qquad (1)$$

for $k(t)$ splits into a constant part $0.18 \leq \bar{\mu} \leq 0.26$ and a bell-shaped part $\dot{\mu}G$ of the GDP. The latter was neglected so far although it is the endogenous part of economic growth, decreases $w(t)$, and increases the capital coefficient $k/y = \mu G$, i. e., the number of years required for producing $k(t)$ with the same productivity for producing $y(t)$. (1) predicts with

$\mu(1 + G \dot{y}/y) = \bar{\mu}$ already the S-functional trade-off observed since 1960 for the G7 between accumulated wealth $k(t)$ and growth rate $\dot{y}/y$. Its analytic solutions are

$$y = \bar{a} / (1 + e^{(T_a - t)/E} + e^{\beta(\tau - t)}) \qquad (2)$$

and

$$k = \bar{\mu} G y_k \equiv \bar{\mu} G \bar{a} / (1 + e^{(T_a + \Delta T - t)/E} + e^{\beta(\tau + \Delta \tau - t)}) \qquad (3)$$

with halftime $T_a$ = 2040 and amplitude $\bar{a}$ = 75.000 US$ in their value of 1991. The time shifts $\Delta \tau$ and $\Delta T$ are obtained by inserting the solutions into (1). For $t > \tau + 1/\beta$ (2) converges into the "industrial evolution"

$$a = \bar{a} / (1 + e^{(T_a - t)/E}). \qquad (4)$$

The function plotted through the life expectancy data in figure 1 is

$$L = L_o + (\bar{L} - L_o)/(1 + e^{(T_a - \bar{L}/2 - t)/E}). \qquad (5)$$

$E$ connects 3 generations. $L(t)$ is heritable [3]. It is the first quantitative and at that an a priori information on demand. Its increase beyond the minimum of $L_o$ = 30 years for maintaining the population results when human life integrates and averages linearly over the existential conditions given by $a(t)$ [2]. This yields a fixed relation $L(a)$ with the precedence of $\bar{L}/2 = 59$ years shown with the bar between the inflection points in figure 1. It specifies the maximum mean unisex life expectancy believed since antiquity to be 120 years with $\bar{L}$ = 118 years.

So far the data support the theory with directly measured parameters, but valid predictions require a complete set of variables. Since $y(t)$ is used and produced one needs two more variables for demand. Human capacity $h(t)$ is with its reaction time $E$ to changing existential conditions suggested by the data as compatible counterpart of $k(t)$. Industrial experience shows that the required working time $w(t)$ is designed into $k(t)$ according to its operating conditions and the technical state of the art. This hidden relation is the second obstacle to predicting growth, especially since $w(t)$ is standardized across businesses. It decreased from the agrarian maximum of 96 hours per week (16 hours per day except Sundays) beyond the left hand side of figure 2 to the present G7 level of 35 hours per week.

$k(t)$ is designed to organize and amplify working time $w(t)$ in speed, precision, or power for producing the GDP. Then $h(t)$ must organize and amplify annual spare time $s(t)$ for using the GDP and enjoying affluence. $s(t)$ was ignored to date because it has no monetary value, but G7 affluence would be unbearable without it. Physically $s(t)$ is as important as $w(t)$ because only one and the same time is passing by for the annual sum of $s(t)$ and $w(t)$. Mankind has always the free choice of producing more or having more time for enjoying whatever it produces. Using the maximum possible working time $\bar{\varepsilon} \equiv 1\,p.a.$ = 96 hours per week as unit for measuring $w(t)$ and $s(t)$ specifies the universal trade-off

$$s(t) = \bar{\varepsilon} - w(t) \qquad (6)$$

between shared annual spare time, paid working time, and usually unpaid but balanced homework for the general case of a couple in a reproducing household.

Although the costs of educating $h(t)$ and designing $k(t)$ contribute fully to $y(t)$ only their amplifying parts can achieve economic equilibrium. $k(t)$ splits for the G7 into $k_w \cong 0.6\,k$ for production and $k_s = k - k_w$ for housing. The required part $h_s$ of $h(t)$ follows from equilibrium because all other quantities are known. Since health, law and order, and good working conditions are as important for the quality of life as classical consumption one gets with (6) for linear amplification the simplest possible equilibrium condition

$$(\bar{\varepsilon} - w)h_s = y = w k_w \quad (7)$$

between demand and supply. It contains the hidden relation between *w(t)* and *k(t)*:

$$w = \bar{\varepsilon} / (1 + k_w / h_s) . \quad (8)$$

The extension of (7) for including the initial agricultural state where $w \to \bar{\varepsilon}$ is given in [4]. The simplest form

$$1/y = 1/\bar{\varepsilon} h_s + 1/\bar{\varepsilon} k_w \quad (9)$$

of the solution for *y(t)* shows that the GDP cannot be larger than the smaller of both inputs. The highest G7 support for $k_w$ of $\bar{\mu}_w = 0.15$ yields $\bar{\varepsilon}\bar{k}_w = \bar{\mu}_w \bar{\varepsilon} G \bar{a} \cong 4\bar{a}$ and $\bar{\varepsilon}\bar{h}_s \cong 1.3\bar{a}$ from (9). Then (8) yields with $w = 0.25\bar{\varepsilon}$ a lowest working time of 24 hours per week. The G7 support their educational level of *h(t)* according to $\bar{h}/E \equiv \bar{v}\bar{a}$ with $\bar{v} \cong 6\%$ of their GDP [4]. Since this yields the educational value $\bar{h} \cong 3.7\bar{a}/\bar{\varepsilon}$ the G7 seem to be on the safe side compared to the required value $\bar{h}_s \cong 1.3\bar{a}/\bar{\varepsilon}$.

When this smaller part of human capacity controls and limits *a(t)* and *L(t)* figures 1 and 2 can be easily understood: after recovery from disaster the G7 developed just along the biologically limited adaptability of mankind to increasing affluence, i. e., to an ever increasing throughput of goods and services per unit of time. Only a biologic limit seems to be strong enough for stabilizing the long term parameters for 8 generations and explaining the economic failure of the largest experiments carried out for reversing the saturating trend of the industrial evolution's growth rate $\dot{a}/a = (1 - a/\bar{a})/E$: the technology push after the former USSR orbited Sputnik in 1957, and the monetary push from 1975 to 2008 for shareholder value and profit. There was no shortage of capital, resources, technology or labor for strong growth in this longest time of peace on G7 soil.

Independent of this explanation the solutions (2) to (6), (8), and (9) are valid predictions with all integrals and derivatives. Besides the national support $\bar{\mu}$ for physical capital the system's relevant parameters are the constants *G*, *E*, and $\bar{L}$. The latter calibrates via $\bar{a}$ the inflationary monetary unit of relevant economic variables with a stable natural unit.